\documentclass[14pt,a4paper]{article}

\usepackage[english]{babel}
\usepackage{amsmath,amssymb,amsfonts,amsthm}
\usepackage[mathscr]{eucal}
\usepackage[all]{xy}
\usepackage{hyperref}
\usepackage{setspace}
\usepackage{upgreek}
\usepackage{hyperref}
\usepackage{cite}

\usepackage{times}
\usepackage{color}

\usepackage{indentfirst}

\begin{document}

\title{\textbf{Resonance and stability \\ of higher derivative theories of derived type}}
\author {D.S.~Kaparulin\footnote{dsc@phys.tsu.ru}$\,\,{}^{1,2}$,\
S.L.~Lyakhovich\footnote{sll@phys.tsu.ru}$\,\,{}^1$\, and
O.D.~Nosyrev\footnote{nosyrev@phys.tsu.ru}$\,\,{}^1$}
\date{\small\textit{
$^1$Physics Faculty, Tomsk State University, Lenina ave. 36, Tomsk 634050, Russia
\\[1mm]
$^2$Lebedev Institute of Physics, Leninsky ave. 53, Moscow 119991,
Russia}}

\maketitle

\begin{abstract}
\noindent We consider the class of higher derivative field equations
whose wave operator is a square of another self-adjoint operator of
lower order. At the free level, the models of this class are shown
to admit a two-parameter series of integrals of motion. The series
includes the canonical energy. Every conserved quantity is unbounded
in this series.  The interactions are included into the equations of
motion such that a selected representative in conserved quantity
series is preserved at the non-linear level. The interactions are
not necessarily Lagrangian, but they admit Hamiltonian form of
dynamics. The theory is stable if the integral of motion is bounded
from below due to the interaction. The motions are finite in the
vicinity of the conserved quantity minimum. The equations of motion
for fluctuations have the derived form with no resonance. The
general constructions are exemplified by the models of the
Pais-Uhlenbeck (PU) oscillator with multiple frequency and Podolsky
electrodynamics. The example is also considered of stable
non-abelian Yang-Mills theory with higher derivatives.
\end{abstract}

\section{Introduction}

In 1950, Pais and Uhlebeck first considered the class of
relativistic higher derivative theories whose wave operator is a
polynomial in the another formally self-adjoint operator of lower
order \cite{PU}. Such models can be termed the theories of derived
type. The general derived model is specified by the constant
coefficient polynomial, termed the characteristic polynomial, and
the lower order operator, being the primary operator. As far as the
wave equations are determined modulo multiplication by a nonzero
constant, only the structure of the characteristic polynomial roots
is relevant. Depending on the structure of roots and primary
operator, the setting of a derived model can describe many
long-known higher derivative field theories, see \cite{AKL18} for
review. The example is the Podolsky electrodynamics \cite{Pod}, the
conformal gravity theories in various dimensions \cite{Pope1,Pope2}
also fall in this class. The extended Chern-Simons  \cite{Deser}
usually serves as prototype of a gauge derived model in the
space-time dimension three. The Pais-Uhlenbeck (PU) oscillator
\cite{PU} is the best known example of the derived-type higher
derivative mechanical model.

The derived theories are often considered in the context of studying
of various aspects of higher derivative dynamics, including the
problem of stability. There is a vast literature on the subject, see
the reviews in \cite{Tomboulis, Smilga, Pope1-1, Pope2-1} and
references therein. The recent studies demonstrate that the higher
derivative theories are not necessary unstable, even though the
canonical energy is unbounded.\footnote{Gauge theories with higher
derivatives can have on-shell bounded energy. } The stability has
been studied from several viewpoints. The non-Hermitian quantum
mechanics is used to construct stable quantum theory of the PU model
\cite{BM,B1,M1}. The same problem is solved by means of alternative
Hamiltonian formulations in \cite{BK,DS}. The stability of classical
paths is studied by numerical simulations and analytical methods in
\cite{Smilga1,Smilga2,Pavsic2,PUPert}, including the estimates of
adiabatic invariants \cite{Boulanger}. The structure of symmetries
and conservation laws of derived type models is studied in
\cite{KLS,KKL}. It is shown that the bounded integrals of motion can
exist in these theories that stabilizes the higher derivative
dynamics.

The conclusions about the stability of higher derivative models are
mostly related to the theories with simple roots of characteristic
polynomial. The models with multiple characteristic polynomial roots
usually demonstrate instability already at free level. The simplest
example of this phenomenon is provided by the fourth order PU
oscillator with multiple frequency. The oscillations resonate, so
the motion is unbounded. The theory with resonance has a bounded
from below integral of motion, though it does not lead to stability
\cite{PUEnergy}. The alternative Hamiltonian formulations for PU
model with resonance are canonically equivalent, and all the
possible Hamiltonians are unbounded \cite{BK}. The multiple root in
the characteristic polynomial has been previously considered as a
indication of instability unless the corresponding degree of freedom
is a pure gauge. As the example of this phenomenon   we mention the
the extended Chern-Simons theory with multiplicity two zero root
which is stable due to the gauge invariance \cite{KKL}.

In papers \cite{KLS16,KKL18,AKL19}, the problem of inclusion of
stable interactions is studied from the viewpoint of consistent
deformation of equations of motion and conserved quantities. It is
demonstrated that the higher derivative theories admit the class of
interactions such that preserve a selected conserved quantity of the
free model. If a bounded from below quantity is conserved at the
interacting level, the dynamics of the non-linear theory is stable.
The interacting equations are non-Lagrangian, but the dynamics is
explicitly covariant and it still admits the Hamiltonian
formulation. For construction of interaction, it is critical that
the free theory is stable. The interaction just preserves the model
stability. In the class of derived theories with resonance at free
level, the dynamics free evolution is unbounded. This means that the
interactions have to stabilize originally unstable theory. The
phenomenon of stabilization of dynamics by the interaction is
well-known in the theories without higher derivatives. For example,
the motions of mechanical system can bounded in the vicinity of the
unstable equilibrium position once the higher order corrections are
accounted for the energy.

In the present paper, we construct the stable interactions for the
class of higher derivative theories of derived type with a single
multiplicity two root of the characteristic polynomial. The wave
operator of general model is given by the square of another formally
self-adjoint operator of lower order. The simplest model is the PU
oscillator with a resonance solutions at free level. We exploit the
idea of stabilization of dynamics by means of interaction.  We
demonstrate that the interactions can be introduced such that the
unbounded energy of the free model becomes bounded from below on the
account of the interactions. The dynamics turns out stable in the
vicinity of the energy minimum just by the reasons of energy
conservation. The equations of motion for fluctuations have derived
type without a resonance. The motions are stable in the range of
energies below a certain level. Above this level, the dynamics
becomes singular. All the proposed stable interactions are
non-Lagrangian, but the equations still admit Hamiltonian
formulation. The theory of small fluctuations in the vicinity of
stationary solution is Hamiltonian with bounded Hamiltonian.

The general construction is illustrated by two examples: the PU
oscillator with multiple frequency, and Podolsky electrodynamics.
For the PU oscillator, we detail the structure of the conserved
quantities at free and interacting level. We notice that the free
theory admits a two-parameter series of conserved quantities. The
canonical energy is included in the series. All the conserved
quantities in the series are unbounded. The particular class of
interactions that preserves a selected representative in the series
of conserved quantities is proposed. If the interaction makes the
integral of motion bounded, the non-linear model turns out stable.
The equations of motion for small fluctuations and the upper bound
are identified for the energy such that the model admits the stable
paths. The Podolsky electrodynamics provides a field-theoretical
example of a similar phenomenon. In this higher derivative theory
with unbounded energy of free model, we find the class of stable
interactions with complex scalar field. The non-abelian
generalization of the proposed interactions are discussed.

The article is organized as follows. In the next section, we
consider the conservation laws of the free PU model with multiple
frequency. We construct a two parameter series of integrals of
motion which includes the canonical energy. The general
representative in the series can be interpreted as the energy of a
certain model without higher derivatives. Proceeding from this
interpretation, we construct the class of stable interactions that
correspond to the (not necessary Lagrangian) theory with higher
derivatives. In section 3, we observe that the theories with bounded
energy are stable in certain sense. We identify all the equilibrium
positions of theory. The equilibrium position at the origin is
unstable, while the additional equilibrium positions can be stable.
The motions are bi-harmonic oscillations in the vicinity of the
stable equilibrium positions. The oscillation frequencies are
different. The estimate for the upper limit of energy is given such
that leads to bounded trajectories. In section 4, we consider the
issue of stability from the viewpoint of Hamiltonian formalism. It
is shown that the interacting theory admits the Hamiltonian
formulation with a bounded from below Hamiltonian. In section 5, we
consider the field theoretical example -- the Podolsky
electrodynamics. In section 6, the further generalizations of the
proposed constructions are discussed. In conclusion, we summarize
the results.

\section{The PU oscillator with a multiple frequency}

Consider the mechanical model with  two coordinates $x(t),y(t)$ and
the action functional
\begin{equation}\label{S-xy}
    S[x(t),y(t)]=\int Ldt\,,\qquad L=m\Big(\dot{x}\dot{y}-\omega^2 (xy-\frac{1}{2}y^2)+
    \frac{1}{2}k(\dot{y}^2-\omega^2y^2)\Big)\,.
\end{equation}
The action includes dimensionless parameter  $k$, while $m$ is a
mass, and $\omega$ is a frequency. The Lagrange equations read
\begin{equation}\label{EL-S-xy-1}
    \frac{\delta S}{\delta x}=-m(\ddot{y}+\omega^2y)=0\,,\qquad
    \frac{\delta S}{\delta y}=-m(\ddot{x}+\omega^2(x-y))-mk(\ddot{y}+\omega^2y)=0\,.
\end{equation}
These equations can be equivalently written as
\begin{equation}\label{ELPUxy}
    \ddot{y}+\omega^2y=0\, ,\qquad \ddot{x}+\omega^2 x =\omega^2y \,.
\end{equation}
Obviously, the parameter $k$ drops out of the equations, while it
cannot be excluded from the action by adding any total derivative to
the Lagrangian. In this sense, the equations (\ref{ELPUxy}) admit a
one-parameter series of Lagrangians (\ref{S-xy}). This simple
observation has consequences: given the series of actions, any
symmetry of the equations leads to the series of the conserved
quantities constructed by Noether theorem. This series is important
for inclusion of stable interactions.

The linear equations (\ref{ELPUxy}) obviously describe the
resonating oscillations: $y$ harmonically oscillates with the
frequency $\omega$, while the oscillator equation for $x$ includes
the same frequency as in the solution for $y$. The latter  is
included as the `` force'' on the r.h.s.  of the equation for $x$
So, the motion in the direction of $x$ is unbounded at free level. A
different interpretation is possible for the same system
(\ref{ELPUxy}): $y$ can be considered as an auxiliary variable
because it can be expressed in terms of $x$ and $\ddot{x}$:
\begin{equation}\label{y-ddx}
         y= x+\frac{\ddot{x}}{\omega^2} + \frac{1}{m\omega^2}\bigg(k\frac{\delta S}{\delta x}-\frac{\delta S}{\delta y}\bigg) \approx x+\frac{\ddot{x}}{\omega^2}\,
    .
\end{equation}
The symbol $\approx$ means the equality on the mass shell. Once the
auxiliary variable $y$ is excluded by means of (\ref{y-ddx}), the
remaining coordinate $x$ obeys a single fourth order equation of the
derived type
\begin{equation}\label{PU-x}
    \ddddot{x}+2\omega^2\ddot{x}+\omega^4x=D^2x=0\,, \qquad D=
    \left(\frac{d^2}{dt^2}+\omega^2\right) \,.
\end{equation}
It is the PU oscillator \cite{PU} with resonance. Equation
(\ref{PU-x}) in itself comes from the action principle with higher
derivatives,
\begin{equation}\label{S-x}
    S_x[x(t)]=\int L_x dt\,,\qquad L_x(x,\dot{x},\ddot{x})=-\frac{1}{2\omega^2}(\ddot{x}^2-2\omega^2\dot{x}^2+\omega^4x^2)\,.
\end{equation}
As we see, the model (\ref{S-xy}) with any $k$ is equivalent to
(\ref{PU-x}) at the level of equations of motion. The solutions,
symmetries and conserved quantities are in one-to-one correspondence
for the models (\ref{S-xy}) and (\ref{S-x}). With this regard, one
can consider (\ref{S-xy}) and (\ref{S-x}) as two different
representations of the same dynamics. The formula (\ref{y-ddx})
expresses $y$ in terms of the derivatives of $x$, $\ddot{x}$, while
$\dot{y}$ admits representation in terms of $\dot{x}$, $\dddot{x}$:
\begin{equation}\label{}
    y=\omega^{-2}\ddot{x}+x\,,\qquad \dot{y}=\omega^{-2}\dddot{x}+\dot{x}\qquad
    \Leftrightarrow\qquad \ddot{x}=\omega^2(y-x)\,,\qquad \dddot{x}=\omega^2(\dot{y}-\dot{x})\,.
\end{equation}
The substitution of $y$ in terms of $x$ and $\ddot{x}$ takes
equations (\ref{EL-S-xy-1}) to (\ref{PU-x}), and vice versa.

We shall seek for the stable interactions such that the equivalence
still exists between the higher derivative representation and the
first order form of this dynamics at the level of equations of
motion. We shall not require, however, the higher derivative
equations to remain Lagrangian upon inclusion of interactions.  The
key role for our construction is played by the series of conserved
quantities parameterized by $k$ involved (\ref{S-xy}). All these
conservation laws are generated by the same symmetry, as we have
already mentioned.

The action functional (\ref{S-xy}), being essentially $k$-dependent,
is invariant with respect to the time translations. The canonical
energy of the model is the conserved quantity,
\begin{equation}\label{PU-xy-Jser}
    \phantom{\frac12}J=m\Big(\dot{x}\dot{y}+\omega^2(xy-\frac{1}{2}y^2)+\frac{1}{2}k(\dot{y}^2+\omega^2y^2)\Big)\,.\phantom{\frac12}
\end{equation}
This expression represents the energy in terms of the coordinates
$x,y$. It explicitly depends on the parameter $k$.  Given the
connection between $y$ and $x$ (\ref{y-ddx}),  the conserved
quantity can be represented in terms of coordinate $x$ and its
higher derivatives
\begin{equation}\label{PU-x-Jser}
    \phantom{\frac12}J=m\Big\{\omega^{-2}\dot{x}\dddot{x}-\frac{1}{2}\omega^{-2}\ddot{x}^2+\dot{x}^2+
    \frac{1}{2}\omega^2x^2+\frac{1}{2}k\Big((\omega^{-2}\dddot{x}+\dot{x})^2+\omega^2(\omega^{-2}\ddot{x}+x)^2\Big)\Big\}\,.\phantom{\frac12}
\end{equation}
As is seen, the constant $k$ is involved in the integral of motion,
while the equation of motion (\ref{PU-x}) is independent of $k$.
This means that $k$-dependent and $k$-independent terms conserve
separately, i.e.
\begin{equation}\label{PU-J12}
    \phantom{\frac12}\frac{d}{dt}\Big(\omega^{-2}\dot{x}\dddot{x}-\frac{1}{2}\omega^{-2}\ddot{x}^2+\dot{x}^2+
    \frac{1}{2}\omega^2x^2\Big)\approx0\,,\qquad
    \phantom{\frac12}\frac{d}{dt}\Big(\omega^{-2}(\dddot{x}+\dot{x})^2+\omega^2(\omega^{-2}\ddot{x}+x)^2\Big)\approx0
    \,.\phantom{\frac12}
\end{equation}
Equations (\ref{PU-xy-Jser}), (\ref{PU-x-Jser}) determine a
two-parameter series of the conserved quantities of the PU theory
with resonance. Each representative in the series can be considered
as the canonical energy of a certain representative of the action
functional series (\ref{S-xy}). This interpretation is not generally
true at the higher derivative formalism (\ref{PU-x}). The canonical
energy of the higher derivative theory (\ref{S-x}) corresponds to
the first entry in the series (\ref{PU-x-Jser}).

The integrals of motion (\ref{PU-J12}) do not result in the
stability of the dynamics. The series (\ref{PU-xy-Jser}) involves
the dynamical variable $x$ in a linear way. Once $x,y$ and
$\dot{x},\dot{y}$ admit independent initial data, the corresponding
conserved quantity is obviously unbounded. As the classical energy
does not have a minimum, one can expect the unbounded  spectrum of
Hamiltonian at quantum level,  and the ``Ostrogradski ghosts"
emerge. Another reason is that the conservation law
(\ref{PU-xy-Jser}) does not restrict the motion to the vicinity of
the critical point $x=y=0$. The general solution of the equations
(\ref{EL-S-xy-1}) read
\begin{equation}\label{xy-sol}
    \phantom{\frac12}
    x(t)=\omega A t \sin (\omega t+a)+B\sin (\omega t+b)\,,\qquad y(t)=2A\cos(\omega t+a)\,,\phantom{\frac12}
\end{equation}
where $A,B$ and $a,b$ are the arbitrary integration constants.  The
amplitude of the first oscillation (\ref{xy-sol}) is linearly
growing with time. The motion is finite only for the special
solutions with $A=0$. The substitution of the general solution
(\ref{xy-sol}) into (\ref{PU-xy-Jser}) gives the expression for the
conserved quantity in terms of initial data:
\begin{equation}\label{J-xy-sol}
    \phantom{\frac12}
    J\Big|_{x=x(t),y=y(t)}=2\omega^2(AB\sin(b-a)+(k-1)A^2)\,.\phantom{\frac12}
\end{equation}
For any fixed value of the parameter $k$, the condition
$J=\text{const}$ does not restrict the amplitudes of oscillations.
The motion is unbounded for any value of $J$. The latter fact
explains the instability of model (\ref{PU-x}) from the viewpoint of
structure of its classical paths. Obviously, the absence of lower
bound of the general representative of the energy series
(\ref{PU-xy-Jser}) is a source of the instability. However, the
stabilization of dynamics is possible upon inclusion of interaction
if the conserved quantity is bounded from below in a vicinity of the
critical point.

At the end of this section, we would like to make a remark on the
role of the parameter $k$ in the series of action functionals
(\ref{S-xy}) and integrals of motion (\ref{PU-xy-Jser}). Obviously,
that the different representatives in these series are connected by
the coordinate transformation,
\begin{equation}\label{xy-k-xy}
    \phantom{\frac12}x\quad\mapsto\quad x'=x+\frac12sy\,,\qquad y\quad\mapsto\quad y'=y\,,\phantom{\frac12}
\end{equation}
where $s$ is the real constant parameter. The transformation with
$s=-k$ brings $k$ to zero in a general representative in
(\ref{S-xy}), (\ref{PU-xy-Jser}). It terms of the higher derivative
description, this transformation does not correspond to any local
change of the coordinate $x$. Instead, we have  a non-local
transformation
\begin{equation}\label{x-k-x}
    \phantom{\frac12}x\quad\mapsto\quad x'=x+\frac12s(\omega^{-2}\ddot{x}+x)\,,\phantom{\frac12}
\end{equation}
which is not invertible in the class of local coordinate changes of
the coordinate $x$. In particular, the variational principle
(\ref{S-xy}) with $k\neq 0$ does not correspond to any variational
principle with higher derivatives and single dynamical variable $x$.
This fact has several consequences at the non-linear level and in
the Hamiltonian formalism, which are discussed in the next two
sections.

\section{Stable interactions}

In this section, we consider the inclusion of interaction in the
Lagrangian (\ref{S-xy}) such that the coordinate $y$ can be still
excluded on shell in terms of $x, \dot{x}, \ddot{x}$, though not
necessarily in the same way as it has been done in the free theory
(\ref{y-ddx}). This would mean that the interactions in the first
order theory (\ref{S-xy}) correspond to the local interactions in
the higher derivative equation (\ref{PU-x}), though the interaction
vertex is not necessarily Lagrangian in the higher derivative setup.

The simplest option is to consider the class of interactions in the
model (\ref{S-xy}) where the coordinate $y$ is expressed from the
same combination of the Lagrangian equations as in the free theory
(\ref{y-ddx}). In the slightly different wording, this means that at
the interacting level, the same linear combination of Lagrangian
equations defines $y$ as function of $x$, $\dot{x}$, $\ddot{x}$:
\begin{equation}\label{y-S-xy}
    y-f(x,\dot{x},\ddot{x})=\frac{1}{m\omega^2}\Big(\frac{\delta S}{\delta y}-k\frac{\delta S}{\delta x}\Big)\,,
\end{equation}
where $f(x,\dot{x},\ddot{x})$ is a function of the coordinate $x$
and its derivatives. This function can be nonlinear, unlike the free
model (\ref{y-ddx}). The most general Lagrangian that has the
property (\ref{y-S-xy}) reads
\begin{equation}\label{S-xy-int}
    L=m\Big(\dot{x}\dot{y}-\omega^2 (xy-\frac{1}{2}y^2)+
    \frac{1}{2}k(\dot{y}^2-\omega^2y^2)+U(x,\dot{x})+V(z,\dot{z})\Big)\,,
\end{equation}
where $U(x,\dot{x})$, $V(z,\dot{z})$ are the functions of the
variables $x,z\equiv x+ky$ and its first time derivatives
$\dot{x},\dot{z}$. The interaction potentials $U$, $V$ are assumed
at least cubic in the variables. It terms of the potentials, the
coordinate $y$ is defined on shell as follows
\begin{equation}\label{y-xdx}
y=f(x,\dot{x},\ddot{x})\equiv\frac{1}{\omega^2}\Big(\ddot{x}+\omega^2x+k\Big(\frac{\partial
U}{\partial x}-\frac{d}{dt}\frac{\partial U}{\partial
    \dot{x}}\Big)\Big)\,.
\end{equation}
The function $V(x,\dot{x})$ does not contribute to this equation.
The higher derivative equation of motion for a single coordinate $x$
has the following form:
\begin{equation}\label{E-x-int}
    \frac{1}{\omega^2}\Big(\frac{d}{dt^2}+\omega^2\Big)^2x+\Big((k-1)+\frac{k}{\omega^2}\frac{d^2}{dt^2}\Big)
    \Big(\frac{\partial U}{\partial x}-\frac{d}{dt}\frac{\partial U}{\partial
    \dot{x}}\Big)-\Big(\frac{\partial V}{\partial z}-\frac{d}{dt}\frac{\partial V}{\partial
    \dot{z}}\Big)\Big|_{z=x+kf(x,\dot{x},\ddot{x})}=0\,,
\end{equation}
where the function $f(x,\dot{x},\ddot{x})$ denotes the right hand
side of equation (\ref{y-xdx}). The obtained equation is the obvious
deformation of the free PU theory equation (\ref{PU-x}). The class
of interacting theories (\ref{S-xy-int}) is not invariant with
respect to the transformation (\ref{xy-k-xy}). This automatically
means that the different representatives of the free action
functional series (\ref{S-xy}) give rise to different classes of
interacting theories, which are unconnected to each other by any
change of coordinates.

Only one representative of the integral of motion series
(\ref{PU-xy-Jser}) conserves at the interacting level. It is given
by the canonical energy of the action functional (\ref{S-xy-int}),
\begin{equation}\label{J-xy}
    J=m\Big\{\dot{x}\dot{y}+\omega^2(xy-\frac{1}{2}y^2)+\frac{1}{2}k(\dot{y}^2+\omega^2y^2)+
    \dot{x}\frac{\partial U}{\partial \dot{x}}-U+
    \dot{z}\frac{\partial V}{\partial \dot{z}}-V\Big\}\,,\quad z=x+ky\,.
\end{equation}
This quantity admits representation in two equivalent forms. First,
the quadratic form of this expression can be brought to the diagonal
form. In so doing, we assume\footnote{ The assumption does not
restrict generality. As we demonstrate below, only interactions with
$k>1$ can be stable.} that $k(k-1)\neq 0 \, $
\begin{equation}\label{J-xy-d}\begin{array}{c}\displaystyle
    J=m\Big\{\frac{k}{2}\Big(\dot{y}+\frac{1}{k}\dot{x}\Big)^2+
    \frac{(k-1)\omega^2}{2}\Big(y+\frac{1}{k-1}x\Big)^2-\frac{1}{2k}\dot{x}^2-\frac{\omega^2}{2(k-1)}x^2+
    \dot{x}\frac{\partial U}{\partial \dot{x}}-U+
    z\frac{\partial V}{\partial \dot{z}}-V\Big\}\,.
\end{array}\end{equation}
Second, the coordinate $y$ can be expressed in terms of $x$ and its
derivatives by means of equation (\ref{y-xdx}). In this way, we
arrive to the conservation law of the higher derivative theory
(\ref{E-x-int}),
\begin{equation}\label{J-x-d}\begin{array}{c}\displaystyle
    J=m\Big\{\frac{k}{2}\Big(\frac{k\omega^{-2}\dddot{x}+(k+1)\dot{x}}{k}+\frac{k}{\omega^2}\frac{d}{dt}\Big(\frac{\partial U}{\partial
    x}-\frac{d}{dt}\frac{\partial U}{\partial\dot{x}}\Big)\Big)^2+
    \frac{(k-1)\omega^2}{2}\Big(\frac{(k-1)\omega^{-2}\ddot{x}+k
    x}{k-1}+
    \frac{k}{\omega^2}\Big(\frac{\partial U}{\partial x}-\\[5mm]\displaystyle-\frac{d}{dt}\frac{\partial
    U}{\partial\dot{x}}\Big)\Big)^2-\frac{1}{2k}\dot{x}^2-\frac{\omega^2}{2(k-1)}x^2+
    \dot{x}\frac{\partial U}{\partial \dot{x}}-U+
    \Big(z\frac{\partial V}{\partial
    \dot{z}}-V\Big)\Big|_{z=x+kf(x,\dot{x},\ddot{x})}\Big\}\,.
\end{array}\end{equation}
As is seen from equations (\ref{J-xy-d}) and (\ref{J-x-d}), the
unbounded contributions come from the terms with squares of $x$ and
$\dot{x}$. The integral of motion of the interacting theory is
bounded if the contributions from the interaction potentials $U$,
$V$ compensate the unbounded terms. The simplest example of
interactions that meet the stability condition reads
\begin{equation}\label{U-xy}
    U(x,\dot{x})=\frac{1}{4\omega^2}\alpha\dot{x}^4+\frac{1}{2}\beta
    x^2\dot{x}^2-\frac{1}{4}\gamma\omega^2x^4\,,\qquad
    V(x,\dot{x},\ddot{x})=0\,,
\end{equation}
where the constants $\alpha,\beta,\gamma$ represent the interaction
parameters.

In this article, we do not explore the dynamics of non-linear model
(\ref{E-x-int}) with  the most general potentials $U,V$ such that
lead to stability. For the sake of technical simplicity, we mostly
focus on the particular class of interaction potentials
(\ref{U-xy}). As we observe below, this class is wide enough to
illustrate the general properties of the dynamics at the non-linear
level. Given the specific potentials,  equation of motion
(\ref{E-x-int}) reads
\begin{equation}\label{E-x-U}\begin{array}{c}\displaystyle
    \Big(1-k(3\alpha\omega^{-2}\dot{x}^2+\beta x^2)\Big)\ddddot{x}-6k
    \Big(3\alpha\omega^{-2}\dot{x}\ddot{x}+\beta x\dot{x}\Big)\dddot{x}+
    \Big(2\omega^2+3\alpha
    ((1-k)\dot{x}^2-6k\omega^{-2}\ddot{x}^2)+\\[5mm]\displaystyle
    +\beta((1-k)\omega^2x^2-k(4x\ddot{x}+7\dot{x}^2))-3\gamma k\omega^2x^2\Big)\ddot{x}+\Big(\beta(1-k)-6\gamma k\Big)\omega^2x\dot{x}^2+
    \omega^4\Big(x-\gamma(k-1)x^3\Big)=0\,.
\end{array}\end{equation}
The integral of motion of the non-linear theory (\ref{E-x-U}) is
given by the expression (\ref{J-x-d}) with the interaction potential
(\ref{U-xy}). The explicit computation gives
\begin{equation}\label{J-x-U}\begin{array}{l}\displaystyle
    J=m\Big\{\frac{k}{2}\Big(\frac{k\omega^{-2}\dddot{x}+(k+1)\dot{x}}{k}-
    \frac{k}{\omega^2}\frac{d}{dt}\Big(3\alpha\omega^{-2}\dot{x}^2\ddot{x}+
    \beta(x^2\ddot{x}+x\dot{x}^2)+\gamma\omega^2x^3\Big)\Big)^2+\\[5mm]\displaystyle\qquad\qquad
    +\frac{(k-1)\omega^2}{2}\Big(\frac{(k-1)\omega^{-2}\ddot{x}+k x}{k-1}-\frac{k}{\omega^2}
    \Big(3\alpha\omega^{-2}\dot{x}^2\ddot{x}+
    \beta(x^2\ddot{x}+x\dot{x}^2)+\gamma\omega^2x^3\Big)\Big)^2-\\[5mm]\displaystyle\qquad\qquad\qquad
    -\frac{1}{2(k-1)}\omega^2x^2-\frac{1}{2k}\dot{x}^2+\frac{3}{4\omega^2}\alpha\dot{x}^4+\frac{1}{2}\beta
    x^2\dot{x}^2+\frac{1}{4}\gamma\omega^2x^4\Big\}\,.
\end{array}\end{equation}
From the viewpoint of stability, the presence or absence of lower
bound of energy is relevant. The first two lines of equation
(\ref{J-x-U}) involve total squares. These terms give positive
contributions to energy if $k>1$ irrespectively to the specifics of
interaction. The third line includes the interaction energy. At free
level it is unbounded. Once the interaction is included, it becomes
bounded if the coupling constants $\alpha,\beta,\gamma$ are
positive.

Depending on the structure of the interaction term in the integral
of motion, the non-linear theory can be stable or unstable. If the
level surfaces of constant energy are bounded in the phase-space,
the motions are finite for all the initial data. It is the case of
globally stable theory. The globally stable interactions are
admitted by the PU oscillator with non-degenerate frequency spectrum
\cite{KL15,KLS16}. For examples of stable interactions in the field
theory we refer articles \cite{AKL18,KKL18,AKL19}. There is no
analogous way to include the globally stable interaction for the PU
oscillator with a multiple frequency. Under the less restrictive
assumptions, the energy can be an unbounded function which admits a
local minimum. Then, the motions are stable in the vicinity of the
energy minimum. This is a case of so-called stability island. The
models with stability island are stable in the range of energies
below certain limit. The highest value of energy with bounded
isoenergetic surface determines the upper energy limit for the
stable paths. In principle, the existence of a stability island is
sufficient for construction of quantum theory with quasi-stable
states and well-defined vacuum state. The precedents are known of
this type for the higher derivative systems \cite{Pavsic2,Smilga},
though not with a resonance at free level. We see that the concept
of an island of stability also suits well for the dynamics of
interacting theory (\ref{E-x-U}). The free model (\ref{PU-x}) is
unstable. To get the stability in the interacting theory, the model
should have a (local) minimum of energy due to interaction. The
motions are bounded in the vicinity of the stable equilibrium
position just by virtue of the energy conservation law.

Depending on the value of the interaction parameters
$\alpha,\beta,\gamma$, the theory (\ref{E-x-U}) may have one or
three stationary solutions. The case $\gamma(k-1)\geq0$ is not
interesting because the model (\ref{E-x-U}) has a single unstable
equilibrium position at the origin. If $\gamma(k-1)>0$, we have two
nonzero stationary solutions:
\begin{equation}\label{Equil-x}
x=\pm \frac{1}{\sqrt{\gamma(k-1)}}\,.
\end{equation}
They can be stable or unstable depending on the values of the
interaction parameters. Introduce the special notation for the
fluctuation in the vicinity of the equilibrium position,
\begin{equation}\label{}
u=x\mp\frac{1}{\sqrt{\gamma(k-1)}}\,.
\end{equation}
The decomposition of the integral of motion (\ref{J-x-U}) in the
vicinity of nonzero stationary solution reads
\begin{equation}\label{J-x-equil}\begin{array}{c}\displaystyle
    J=m\Big\{-\frac{\omega^2}{4\gamma(k-1)^2}+\frac{1}{2k}\Big(\frac{1}{\omega^2}\Big(1-\frac{\beta k}{\gamma(k-1)}\Big)
    \dddot{u}-\frac{2k^2+1}{k-1}\dot{u}\Big)^2+\\[5mm]\displaystyle
    +\frac{\omega^2}{2(k-1)}\Big(\frac{1}{\omega^2}\Big(1-\frac{\beta k}{\gamma(k-1)}\Big)
    \ddot{u}-\frac{2k}{k-1}u\Big)^2+\frac{1}{2k}\Big(\frac{\beta k}{\gamma(k-1)}-1\Big)
    \dot{u}^2+\frac{\omega^2}{k-1}u^2+\ldots\Big\}\,,
\end{array}\end{equation}
The dots denote cubic and higher terms in $u$ and its derivatives.
The first term defines the value of energy at the stationary
solution. We introduce the special notation for this value,
\begin{equation}\label{Jmin}
    J_{min}=-\frac{m\omega^2}{4\gamma(k-1)^2}\,.
\end{equation}
This value is negative for $\gamma>0,k>1$. The stability properties
of the equilibrium position are determined by signature of the
quadratic form in the decomposition (\ref{J-x-equil}). The energy
has minimum if all the coefficients at squares are positive,
\begin{equation}\label{abc-cond}
    k-1>0\,,\qquad \gamma>0\,,\qquad \beta k - \gamma (k-1)>0\,.
\end{equation}
Hereinafter, it is assumed that the stability conditions are
satisfied. In particular, we suppose below that $\beta,\gamma>0$.
The parameter $\alpha$ is not involved in the stability conditions
because the $\dot{x}^4$ term cannot influence the motions in the
vicinity of the equilibrium position. However, this term influences
the stability properties of the model at higher energies.

The model (\ref{E-x-U}) cannot be globally stable. The stable
interactions imply the positive values of the interaction parameters
$\beta,\gamma$. Once this parameters are positive, the coefficient
at the highest derivative term in the equation of motion can vanish,
\begin{equation}\label{zero-H}
        \phantom{\frac12}1-k(3\alpha\omega^{-2}\dot{x}^2+\beta x^2)=0\,.\phantom{\frac12}
\end{equation}
On this phase-space surface, the conditions of existence and
uniqueness of solution to the Cauchy problem for equation
(\ref{E-x-U}) are violated. Our analysis shows that relations
(\ref{E-x-U}) and (\ref{zero-H}) are inconsistent. This means that
neither true trajectory can be transverse to this surface, no the
classical path can lie on the phase-space submanifold
(\ref{zero-H}). The only alternative is that true trajectories begin
or end in the vicinity of the surface (\ref{zero-H}). In the
globally stable theories, the classical path are
complete.\footnote{The path $x(t)$ is complete if it is defined for
the real value of time $-\infty<t<+\infty$.} It is not possible for
the system (\ref{E-x-U}) as we see. The complete trajectories can
exist if the certain level energy surface defined by (\ref{J-x-U})
has no intersection with the phase-space submanifold (\ref{zero-H}).
The stationary solutions (\ref{Equil-x}) are examples of complete
paths. They do not lie on the surface (\ref{zero-H}) because
conditions (\ref{Equil-x}) and (\ref{zero-H}) are contradictory. The
classical paths in the vicinity of stable equilibrium position have
a good chance to be complete. The level surfaces of integral of
motion (\ref{J-x-U}) are compact and lie in the neighborhood of the
equilibrium position with no intersection with the phase-space
submanifold (\ref{zero-H}) if the energy value is sufficiently
small. The biggest compact isoenergetic surface (\ref{J-x-U}) that
has no intersection with the phase-space submanifold (\ref{zero-H})
determines the border of stability island. The brief summary of the
above is the following. The stable paths of the system (\ref{E-x-U})
are localized in two islands of stability, which lie in the vicinity
of nonzero stationary solutions (\ref{Equil-x}). The existence of
singular surface (\ref{zero-H}) is the obstruction to the global
stability for the class of interactions (\ref{E-x-U}).

Consider the dynamics of small fluctuations in the vicinity of the
nonzero stationary solution (\ref{Equil-x}). The linearization of
equation (\ref{E-x-int}) in the vicinity of solution (\ref{Equil-x})
reads
\begin{equation}\label{E-x-equil}
    \Big(\frac{\beta k}{\gamma(k-1)}-1\Big)\ddddot{u}+
    \Big(\frac{\beta}{\gamma}+\frac{k+2}{k-1}\Big)\omega^2\ddot{u}+2\omega^4u=0\,,
\end{equation}
where $u$ is the fluctuation. By construction, the equation of
motion has the PU form. The frequencies of oscillations are
determined by the interaction parameters and the constant $k$. The
stability conditions (\ref{abc-cond}) imply that all the
coefficients of equation are positive, so the roots of
characteristic equations are complex. The frequencies of
oscillations for the model (\ref{E-x-equil}) read
\begin{equation}
 \omega_{\pm}=\omega\sqrt{\frac{\displaystyle\beta (k-1)+
 \gamma(k+2)\pm\sqrt{(\beta (k-1)+
 \gamma(k+2))^{2}-
 8\gamma(k-1)(\beta k-(k-1)\gamma)}}{\displaystyle
 2(k\beta-\gamma(k-1))}}\,.
\end{equation}
The conditions (\ref{abc-cond}) imply that the frequencies are not
equal, so the system (\ref{E-x-equil}) has no resonance at the
equilibrium position. A simplest way to see the fact is follows. The
equations (\ref{E-x-int}) admit alternative formulation without
higher derivatives (\ref{S-xy-int}). The resonance has place if both
the oscillators one and the same frequency. This option is not
possible because the system of two free oscillators with one and the
same frequency does not allow reformulation in terms of higher
derivative PU theory with a single dynamical variable. The solution
to the equation of motion is bi-harmonic oscillation,
\begin{equation}\label{}
    u=A\sin(\omega_+t+a)+B\sin(\omega_-t+b)\,,
\end{equation}
where $A,B,a,b$ are integration constants. The bi-harmonic
oscillation is a finite path, so the dynamics should be considered
as stable. The account of interaction does not change conclusion of
about the stability of motion. The argument is that the dynamics is
localized on zero energy surface, which is compact for the energies
slightly above the lower bound. All this means that the dynamics of
non-linear theory is stable in the vicinity of equilibrium position.

Let us now specify the stability island. The classical path is
singular if the surfaces (\ref{J-x-U}) and (\ref{zero-H}) are
intersect. For regular path conditions (\ref{J-x-U}) and
(\ref{zero-H}) are inconsistent. The regularity condition is met in
the range of energies
\begin{equation}\label{Jmin-Jmax}
    J_{min}\leq J< J_{max}\,,
\end{equation}
where $J_{min}$ is the minimal value of energy (\ref{Jmin}), and $J_{max}$ is the minimal value of
energy on the singular surface (\ref{zero-H}). Expressing the coordinate $x$ on the surface
(\ref{zero-H}), we represent the energy  (\ref{J-x-U}) as the function of
three variables,
\begin{equation}\label{J-zero-H}\begin{array}{c}\displaystyle
    J\bigg[x\mapsto\pm\sqrt{\frac{1}{\beta}\Big(\frac{1}{k}-\frac{3\alpha}{\omega^2}\dot{x}^2\Big)}\bigg]=
    m\bigg\{\frac{k}{2}\bigg[\Big(\frac{k+1}{k}-\frac{3\gamma}{\beta}\Big)\dot{x}-
    \frac{6\alpha k\dot{x}\ddot{x}^2}{\omega^4}\mp\frac{4\dot{x}\ddot{x}}{\omega^2}\sqrt{\beta\Big(\frac{1}{k}-\frac{3\alpha}{\omega^2}\dot{x}^2\Big)}
    +\\[5mm]\displaystyle+\frac{k}{\omega^2}\Big(\frac{9\gamma\alpha}{\beta}-\beta\Big)\dot{x}^3\bigg]^2+\frac{k^3\omega^2}{2\beta(k-1)}\Big(\frac{1}{k}-\frac{3\alpha}{\omega^2}\dot{x}^2\Big)
    \bigg[\frac{1}{k-1}-\sqrt{\beta\Big(\frac{1}{k}-\frac{3\alpha}{\omega^2}\dot{x}^2\Big)}-\frac{\gamma}{\beta}\Big(\frac{1}{k}-\frac{3\alpha}{\omega^2}\dot{x}^2\Big)\bigg]^2+\\[5mm]\displaystyle+\frac{3\alpha}{4\omega^2}
    \Big(\frac{3\gamma\alpha}{\beta^2}-1\Big)\dot{x}^4+\frac{3\alpha}{2\beta(k-1)}\Big(1-\frac{\gamma(k-1)}{\beta k}\Big)\dot{x}^2
    -\frac{\gamma\omega^2}{4k^2\beta^2}\Big(\frac{3\beta k}{\gamma(k-1)}-1\Big)
    \bigg\}\,.
\end{array}\end{equation}
The domain of a function is $|\dot{x}|<\sqrt{1/3k\alpha}$ for
$\alpha>0$, and all the phase space for $\alpha\leq0$. Otherwise the
condition (\ref{zero-H}) is inconsistent. The arguments in squares
in two first lines of (\ref{J-zero-H}) are independent initial data,
which account the dependence of the energy on $\ddot{x},\dddot{x}$.
The value of these terms can be set zero irrespectively to the
$x,\dot{x}$ terms of third line. The actual minimum of the function
is given by minimum of the quadratic form in $\dot{x}^2$ in the
third line of the expression (\ref{J-zero-H}). Three different cases
are summarized in the equation below,
\begin{equation}\label{Jmax}
    J_{max}=\left\{%
\begin{array}{l}\displaystyle
    -\frac{m\gamma\omega^2}{4\beta^2k^2}\bigg(\frac{2\beta k}{\gamma(k-1)}-1\bigg)\,,\\[5mm]\displaystyle
    \qquad\alpha=0\qquad\text{or}\qquad\alpha\geq\frac{\beta^2}{3\gamma}\qquad\text{or}\qquad
    0<\alpha<\frac{\beta^2}{3\gamma},\qquad \frac{\beta(\beta k-\gamma(k-1))}
    {k(k-1)(\beta^2-3\gamma\alpha)}\geq\frac{1}{3\alpha k};\\[5mm]\displaystyle
    -m\omega^2\Big\{\frac{\gamma}{4\beta^2k^2}\Big(\frac{2\beta k}{\gamma(k-1)}-1\Big)
    -\frac{1}{2\beta(k-1)}\Big(1-\frac{\gamma(k-1)}{\beta k}\bigg)-\frac
    {1}{4}
    \Big(\frac{3\gamma\alpha}{\beta^2}-1\Big)\Big\}
    \\[5mm]\displaystyle
    \qquad 0<\alpha<\frac{\beta^2}{3\gamma},\qquad \frac{\beta(\beta k-\gamma(k-1))}
    {k(k-1)(\beta^2-3\gamma\alpha)}<\frac{1}{3\alpha k};\\[5mm]\displaystyle
    -\frac{m\gamma\omega^2}{4\beta^2k^2}\Big(\frac{2\beta k}{\gamma(k-1)}-1\Big)-
    \frac{3m|\alpha|\omega^2}{4k(k-1)^2}\frac{k\beta-\gamma(k-1)}{3\gamma|\alpha|+\beta^2}\,,
    \\[5mm]\displaystyle
    \qquad\alpha<0\,.
\end{array}%
\right.
\end{equation}
The estimates in the right hand side of this equation give the upper
bound for the energy of the regular path. The surface $J=J_{max}$ is
the border of the stability island.

Let us summarize the results of the section. If the parameters of
the model meet the condition (\ref{abc-cond}), the non-linear theory
(\ref{E-x-U}) has three equilibrium positions. The equilibrium
position at the origin, being the critical point of the free theory,
is unstable. The other two equilibrium points are stable. The
motions are regular for energies in the range (\ref{Jmin-Jmax}),
where $J_{min}$, $J_{max}$ are given by relations (\ref{Jmin}),
(\ref{Jmax}). The maximal energy of the regular trajectories defines
the boundary of the stability island. The value $\alpha=0$
corresponds to the maximal possible size of the stability island
(the first option in equation (\ref{Jmax})). The size of the
stability island decreases for negative $\alpha$, which contributes
a negative correction to the kinetic term, and also for small
positive $\alpha$. The latter decrease of the energy limit may seem
contrintuitive. We explain it by almost degeneracy of the energy
quadratic form (\ref{J-x-equil}) at the equilibrium position (the
coefficient at $\dot{u}$ is small). In this case, every small change
of the model parameters can have a negative impact on stability.

\section{Hamiltonian formalism}

Let us begin with the free PU theory. At first, consider the
canonical Hamiltonian formalism for the theory with the action
functional (\ref{S-xy}). The action involves the parameter $k$,
which is not involved in the equivalent higher derivative theory
(\ref{S-x}), so we aim at clarifying the consequences of this
ambiguity in Hamilitonian formalism.

Introduce the canonical momenta $p_x$, $p_y$ obeying the canonical
\begin{equation}\label{pxpy}
p_x=m\dot{y}\,,\qquad p_y=m(\dot{x}+k\dot{y})\, ,
\end{equation}
\begin{equation}\label{PB-can}
   \phantom{frac12}\{x,p_x\}=\{y,p_y\}=1\,,\qquad
   \{x,y\}=\{p_x,p_y\}=\{x,p_y\}=\{y,p_x\}=0\,.\phantom{frac12}.
\end{equation}
The Hamiltonian equations read
\begin{equation}\label{xy-Ham}
    \dot{x}=\{x,H\}\,,\qquad \dot{p}_x=\{p_x,H\}\,,
    \qquad \dot{y}=\{y,H\}\,,\qquad \dot{p}_y=\{p_y,H\}\,,
\end{equation}
\begin{equation}\label{xy-H}
    H=\frac{1}{m}(p_xp_y+\frac{1}{2}kp_x^2)+m\omega^{2}(xy+\frac{k-1}{2}y^2)\,.
\end{equation}
The equivalence between the Hamiltonian equations (\ref{xy-Ham}),
(\ref{xy-H}) and higher derivative PU theory (\ref{PU-x}) is easy to
see: The momenta $p_x,p_y$ and auxiliary variable $y$ can be
expressed in terms of the coordinate $x$ and its derivatives. The
remaining dynamical variable $x$ obeys the higher derivative PU
equation (\ref{PU-x}).

The  Hamiltonians (\ref{xy-Ham}) depend on the parameter $k$. This
dependence is due to the canonical transformation,
\begin{equation}\label{CT}\begin{array}{c}\displaystyle
    x\mapsto x'=x+\frac{1}{2}sy\,,\qquad p_x\mapsto (p_x)'=p_x-\frac{1}{2}sp_y\,,\qquad y\mapsto y'=y\,,\qquad
    p_y\mapsto (p_y)'=p_y\,,
\end{array}\end{equation}
where $s$ is the transformation parameter, being real number. The
Ostrogradski canonical formulation \cite{Ostr} corresponds the
choice $k=0$. The transformation (\ref{CT}) with $s=k$ connects the
general Hamiltonian (\ref{xy-H}) with the canonical Hamiltonian by
Ostrogradski. The subtlety is that the Hamiltonian formulations with
$k\neq0$ cannot be derived by means of the Legendre transformation
of any higher derivative action functional. The paper \cite{GLT}
tells us that all the Hamiltonian formulations, being obtained by
different ways of reduction of order in the Lagrangian, are
connected by the canonical transformation, which does not involve
the original coordinates. This is not the case at hands. The
transformation (\ref{xy-k-xy}) of single dynamical $x$ is not a
change of the coordinate (\ref{x-k-x}). It is some kind of hidden
symmetry, being connected to the resonance. At the interacting
level, the transformation (\ref{xy-k-xy}) no longer preserves the
equations of motion (\ref{E-x-U}). In this case, the Legendre
transformation of the action (\ref{S-xy-int}) allows to construct
the Hamiltonian formulations, which do not follow from the
Ostrogradksi procedure for the higher derivative action (\ref{S-x}).
This Hamiltonian formulations can have bounded Hamiltonian even if
the higher derivatives are involved at the free and interacting
levels. That is why the Hamiltonian formulations (\ref{xy-Ham})
,(\ref{xy-H}) are useful from the viewpoint of inclusion of
interactions.

The Hamiltonian formulations (\ref{xy-Ham}), (\ref{xy-H}) can be
rewritten in terms of the phase-space variables
$x,\dot{x},\ddot{x},\dddot{x}$, being derivatives of $x$. The
transformation law reads
\begin{equation}\label{}
    \dot{x}=\frac{1}{m}(p_y-kp_x),,\qquad \ddot{x}=\omega^2(x-y)\,,\qquad \dddot{x}=\frac{1}{m}\omega^2((k+1)p_x-p_y)\,.
\end{equation}
The Hamiltonian (\ref{xy-H}) is given by the general representative of the
conserved series (\ref{PU-x-Jser}),
\begin{equation}\label{x-H}
    H=m\Big\{\omega^{-2}\dot{x}\dddot{x}-\frac{1}{2}\omega^{-2}\ddot{x}^2+\dot{x}^2+
    \frac{1}{2}\omega^2x^2+\frac{1}{2}k\Big((\omega^{-2}\dddot{x}+\dot{x})^2+\omega^2
    (\omega^{-2}\ddot{x}+x)^2\Big)\Big\},.
\end{equation}
The Poisson brackets of the phase-space variables $x,\dot{x},\ddot{x},\dddot{x}$ read
\begin{equation}\label{PB}
    \{x,\dot{x}\}=-\frac{1}{m}k,\qquad \{\dot{x},\ddot{x}\}=\{\dddot{x},x\}=\frac{1}{m}\omega^2(k+1),\qquad
    \{\ddot{x},\dddot{x}\}=\frac{1}{m}\omega^4(k+2)\,,\qquad\{x,\ddot{x}\}=\{\dot{x},\dddot{x}\}=0\,.
\end{equation}
The Hamiltonian formulation (\ref{x-H}), (\ref{PB}) has been first
proposed in \cite{BK} for the PU oscillator. The canonical
equivalence of different Hamiltonian formulations has been also
noticed in this article. The special feature of the Poisson bracket
(\ref{PB}) is that the coordinate $x$ and its velocity $\dot{x}$ are
conjugated if $k\neq0$. In the Ostrogradski construction of the
Hamiltonian formalism, the coordinate always Poisson commute to the
velocity. This observation means the Hamiltonian formulation with
$k\neq 0$ cannot be derived by the Legendre transformation of a
higher derivative Lagrangian with a single dynamical variable $x$.

Let us now focus on the Hamiltonian formulation for the higher
derivative  equation with interaction (\ref{U-xy}). Even though the
interaction vertices are non-Lagrangian, the equivalent lower
formulation admits the action principle (\ref{S-xy-int}). So, one
can proceed from the action (\ref{S-xy-int}). Introduce the
canonical momenta
\begin{equation}\label{}
    p_x=m(\dot{y}+\alpha\omega^{-2}\dot{x}^3+\beta x^2\dot{x})\,,\qquad p_y=m(\dot{x}+k\dot{y})\,.
\end{equation}
Form these equations the velocity $\dot{y}$ can be expressed in terms of $p_x,p_y$
and $\dot{x}$, while $\dot{x}$ is determined as a solution of algebraic equation,
\begin{equation}\label{dx-eq}
\phantom{\frac12}\dot{y}=\frac{1}{m}p_x-\alpha\omega^{-2}\dot{x}^3-\beta
x^2\dot{x}\,,\qquad k\alpha\dot{x}^3+(k\beta
x^2-1)\dot{x}+\frac{1}{m}(p_y-kp_x)=0\,.\phantom{\frac12}
\end{equation}
This equation has third order if $\alpha\neq0$, otherwise it is
linear. The cubic equation can be solved with respect to $\dot{x}$
by several methods, for example, Cardano formula. Choosing the
appropriate branch of the solution of equation (\ref{dx-eq}), and
computing the Hamiltonian, one can construct the Hamiltonian
formulation for the interacting model (\ref{E-x-U}). It should be
outlined that for stable interactions the relative signs of the
cubic and linear in $\dot{x}$ terms can be opposite. In this case
none, of the branches of the solutions of the cubic equation is
globally defined. This obstruction prevents the construction of
globally defined Hamiltonian formalism in the theory (\ref{E-x-U}).
The number of solution branches in changed on the singular surface
(\ref{zero-H}). If $\alpha=0$, the velocity is explicitly expressed
from the equation (\ref{dx-eq}). The expression for the Hamiltonian
is given below,
\begin{equation}\label{}\begin{array}{c}\displaystyle
    H=\frac{\beta x^2}{2m}\Big(\frac{p_y-kp_x}{\beta k x^2-1}\Big)^2+\frac{1}{2m}
    \frac{(kp_x+(\beta k x^2-2)p_y)(\beta x^2 p_y-p_x)}{(\beta k x^2-1)^2}+m\omega^2\Big(xy-\frac{k-1}{2}y^2+\frac{1}{4}\gamma x^4\Big)\,.
\end{array}\end{equation}
The Hamiltonian is singular at the surface (\ref{zero-H}). So, the
Hamiltonian dynamics is not smooth in the vicinity of the singular
surface even in the case $\alpha=0$.

Let us consider Hamiltonian description of the dynamics of small
fluctuations in the vicinity of the stable equilibrium position. The
decomposition of the Lagrangian (\ref{S-xy-int}), (\ref{U-xy}) in
the vicinity of stationary solution (\ref{Equil-x}) reads
\begin{equation}\label{}\begin{array}{c}\displaystyle
    S[u(t),v(t)]=\int L dt,\\[5mm]\displaystyle
    L=m\Big\{\frac{\omega^2}{4\gamma(k-1)^2}+\frac{\beta}{2\gamma(k-1)}\dot{u}^2+\dot{u}\dot{v}+\frac{1}{2k}\dot{v}^2-
    \omega^2\Big(\frac{3}{2(k-1)}u^2+uv+\frac{1}{2}(k-1)v^2\Big)+\ldots\Big\}\,.
\end{array}\end{equation} The dynamical variables are fluctuations in the
vicinity of nonzero equilibrium position (\ref{Equil-x}),
\begin{equation}\label{}
    u=x\mp\frac{1}{\sqrt{\gamma(k-1)}}\,,\qquad
    v=y\pm\frac{1}{(k-1)\sqrt{\gamma(k-1)}}\,.
\end{equation}
The dots denote the cubic and higher terms in $u,v$. The generalized
momenta $p_u,p_v$ are defined as follows:
\begin{equation}\label{pupv}
    p_u=m\Big(\frac{\beta}{\gamma(k-1)}\dot{u}+\dot{v}\Big)\,,\qquad p_v=m(\dot{u}+k\dot{v}).
\end{equation}
The generalized momenta $\dot{u},\dot{v}$ are expressed as follows:
\begin{equation}\label{dudv}
    \dot{u}=\frac{1}{m}\frac{\gamma(k-1)(p_v-kp_u)}{\gamma(k-1)-\beta k}\,,\qquad \dot{v}=\frac{1}{m}\frac{\gamma(k-1)p_u-\beta p_v}{\gamma(k-1)-\beta k}.
\end{equation}
The Hamiltonian of the model reads
\begin{equation}\label{Huv}
    H=-\frac{m\omega^2}{4\gamma(k-1)^2}+\frac{1}{m}\frac{\gamma(k-1)}{\gamma(k-1)-\beta k}
    \Big(\frac12kp_u^2-p_up_v+\frac{1}{2}\frac{\beta}{\gamma(k-1)}p_v^2\Big)+
    m\omega^2\Big(\frac{3}{2(k-1)}u^2+uv+\frac{1}{2}(k-1)v^2\Big)\,.
\end{equation}
The Hamiltonian is bounded from below if the conditions
(\ref{abc-cond}) are met. This means that the theory of small
fluctuations is stable at the classical and quantum level.

The canonical coordinates $v,p_u, p_v$ are expressed in terms the
derivatives of the fluctuation $u$ and its derivatives up to the
third order,
\begin{equation}\label{}\begin{array}{c}\displaystyle
    v=\frac{1}{\omega^2}\Big(1-\frac{\beta k}{\gamma(k-1)}\Big)\ddot{u}-\frac{2k+1}{k-1}u\,,\qquad
    p_u=\frac{1}{\omega^2}\Big(1-\frac{\beta
    k}{\gamma(k-1)}\Big)\dddot{u}+\Big(\frac{\beta}{\gamma(k-1)}-\frac{2k+1}{k-1}\Big)\dot{u}\,,\\[5mm]\displaystyle
    p_v=\frac{k}{\omega^2}\Big(1-\frac{\beta
    k}{\gamma(k-1)}\Big)\dddot{u}-\frac{2k^2+1}{k-1}\dot{u}\,.
\end{array}\end{equation}
The inverse transformation reads (expression for $\dot{u}$ see in
(\ref{dudv}))
\begin{equation}\label{}
    \ddot{u}=\omega^2\frac{\gamma(k-1)v+\gamma(2k+1)u}{\gamma(k-1)-\beta
    k}\,,\qquad
    \dddot{u}=\Big(\frac{\gamma(k-1)}{\gamma(k-1)-\beta
    k}\Big)^2\Big(\Big(\frac{2k+1}{k-1}-\frac{\beta}{\gamma(k-1)}\Big)p_v-
    \frac{2k^2+1}{k-1} p_u\Big)\,.
\end{equation}
In terms of derivatives of $u$, the Hamiltonian (\ref{Huv}) takes
the form of integral of motion (\ref{J-x-equil}). The Poisson
bracket reads
\begin{equation}\label{PB-1}\begin{array}{c}\displaystyle
    \{u,\ddot{u}\}=\{\dot{u},\dddot{u}\}=0\,,\quad \{u,\dot{u}\}=-\frac{1}{m}\frac{\gamma k (k-1)}{\gamma(k-1)-\beta k},
    \quad \{u,\dddot{u}\}=\{\ddot{u},\dot{u}\}=\frac{1}{m}\omega^2\Big(\frac{\gamma(k-1)}{\gamma(k-1)-\beta k}\Big)^2\frac{2k^2+1}{k-1},\\[5mm]\displaystyle
    \{\ddot{u},\dddot{u}\}=-\frac{1}{m}\omega^4\Big(\frac{\gamma(k-1)}{\gamma(k-1)-\beta
    k}\Big)^3\Big(\frac{\beta}{\gamma(k-1)}+\frac{3}{k-1}\Big)\,.
\end{array}\end{equation}
The coordinate $u$ and velocity $\dot{u}$ are inevitably have
nonzero Poisson bracket for stable interactions. This automatically
means that such Hamiltonian does not follow from the Ostrogradki
procedure. This is not surprising because the Ostrogradski
Hamiltonian of a non-singular higher derivative theory is not
bounded from below, while the function (\ref{Huv}) is bounded.

The example of the higher derivative PU theory (\ref{E-x-U}) with
the multiple frequency tells us that the stable vertices are
possible, though they are not necessarily Lagrangian. One more
conclusion is that the higher derivative equations with
non-Lagrangian interactions can admit Hamiltonian formulation,
though it is inequivalent to any Ostrogradski formalism. If the
quantization of the theory of fluctuations (\ref{E-x-equil}) is
constructed by the means Hamiltonian formulation (\ref{pupv}),
(\ref{Huv}), the classical stubility will persits at quantum level.
Being equivalent to the system of two oscillators, this model has
the usual equidistant spectrum of energy, and it admits the well
defined vacuum state. The unharmonic terms can be accounted for by
perturbation theory. As the wave functions of stationary states of
harmonic oscillator exponentially decreasing at infinity, the
perturbation theory is well defined in each order. In principle,
this is sufficient for perturbative construction of the stable
quantum theory of the non-linear model (\ref{E-x-U}).

\section{A higher derivative field-theory with the resonance}

The Podolsky electrodynamics without Maxwell term provides a
simplest example gauge field theory  with resonance. The action
reads\footnote{We use mostly negative convention for the Minkowski
metrics.}
\begin{equation}\label{Pod-free}
    S[A(x)]=\frac{1}{4}\int \partial_\nu F^{\nu\mu}\partial^\rho F_{\rho\mu}d^dx\,,
\end{equation}
where $F_{\mu\nu}=\partial_\mu A_\nu-\partial_\nu A_\mu$ is the
field strength. In the general Podolsky model, the action
 also includes the Maxwell term, so one of the
photons is massless, while another one is massive. The mass spectrum
of the model  (\ref{Pod-free}) is degenerate: both the
subrepresentations are massless.  To our knowledge, this theory has
not been studied in the literature yet. Maybe it does not attract
the interest because the representation with degenerate mass
spectrum is non-unitary. We view (\ref{Pod-free}) as a toy model
that exemplifies the stability issue in higher derivative field
theory with the resonance at free level, leaving aside the
interpretation of the non-unitary representation.

Similar to the PU model (\ref{S-x}), the theory (\ref{Pod-free})
admits an equivalent formulation without higher derivatives. The
analog of the lower derivative action (\ref{S-xy}) reads
\begin{equation}\label{AB-Pod}
    \phantom{\frac12}\mathcal{L}=-\frac{1}{4}kG_{\mu\nu}G^{\mu\nu}-\frac{1}{2}G_{\mu\nu}F^{\mu\nu}+
    \frac{1}{2}m^2B_{\mu}B^\nu\,,\phantom{\frac12}
\end{equation}
where $G_{\mu\nu}=\partial_\mu B_\nu-\partial_\nu B_\mu$, and $m$ is the constant with mass dimension. We introduce $m$ for the
reasons of convenience. The dynamical variables are the vector fields $A_\mu(x),B_\mu(x)$. The Lagrange equations read
\begin{equation}\label{}
    \frac{\delta S}{\delta B^\mu}=\partial^\mu (F_{\mu\nu}+kG_{\mu\nu})+m^2B_{\mu}=0\,,\qquad
    \frac{\delta S}{\delta A_\mu}=\partial_\nu G^{\mu\nu}=0\,.
\end{equation}
The first equations defines the vector field $B_\mu$ in terms of
derivatives of $A$, $B_\mu\approx-m^{-2}\partial^{\nu}F_{\mu\nu}$.
Then, the second equation means that $A$ should obey the "double
massless" Podolsky equation $\Box\partial^\nu F_{\nu\mu}=0$.
Obviously, the mass $m$ is an accessory parameter that does not
contribute to the equations, much like the parameter $k$ in the
first order equivalent (\ref{S-xy}) of the  PU action (\ref{S-x}).

Much like the mechanical analogue (\ref{PU-xy-Jser}) of the previous
section, the first-order theory (\ref{AB-Pod}) admits a
two-parameter series of conserved tensors.  So, the higher
derivative equivalent (\ref{Pod-free}) should also admit the series
of conserved quantities.
\begin{equation}\label{Pod-AB-Jser}
    \Theta^{\mu\nu}=G^{\mu}{}_\rho F^{\nu\rho}+G^{\nu}{}_\rho F^{\mu\rho}-
    \frac{1}{2}\eta^{\mu\nu}G_{\rho\sigma}F^{\rho\sigma}+k(G^{\mu}{}_{\rho}G^{\nu\rho}-\frac{1}{4}\eta^{\mu\nu}G_{\rho\sigma}G^{\rho\sigma})+
    m^2(B^\mu B^\nu-\frac{1}{2}\eta^{\mu\nu}B_\rho B^\rho)\,.\phantom{\frac12}
\end{equation}
The parameter of the series is the real number $k$. The canonical
energy-momentum tensor of the higher derivative theory
(\ref{Pod-free}) is included in the series (\ref{Pod-AB-Jser}) for
$k=0$. The general representative of the series (\ref{Pod-AB-Jser})
is associated with canonical energy-momentum tensor of the theory
(\ref{AB-Pod}). The energy density is given by the $00$-component of
the conserved tensor. The energy density is unbounded because the
field $A_\mu(x)$ is involved into (\ref{Pod-AB-Jser}) in a linear
way. Because of this observation, the theory (\ref{AB-Pod}) is
unstable at free level. To stabilize the dynamics at the interacting
level, the terms with quadratic dependence on the vector field
$A_\mu(x)$ are needed. To make it in an explicitly covariant way we
include the complex scalar field. From this perspective, we slightly
deviate from the pattern of the mechanical model considered in the
previous section, where no extra degree of freedom is needed for the
stabilization at the interacting level.

The theory (\ref{AB-Pod}) of the vector fields $A_\mu(x)$ and
$B_\mu(x)$ admits the following interactions with the complex scalar
field $\varphi(x)$,
\begin{equation}\label{AB-Pod-U}\begin{array}{c}\displaystyle
    \mathcal{L}=\frac{1}{2}(G_{\mu\nu}F^{\mu\nu}+B_{\mu}B^\nu+\frac{1}{2}kG_{\mu\nu}G^{\mu\nu})+
    \frac{1}{2}(D_\mu\varphi^\ast D^\mu\varphi+m^2(\alpha|\varphi|^2+\beta |\varphi|^4)
    +\frac{1}{4}\gamma|\varphi|^2F_{\mu\nu}F^{\mu\nu}\,,\\[5mm]\displaystyle
    \phantom{\frac12}D_\mu\varphi=(\partial_\mu-ieA_\mu)\varphi\,,\qquad
    D_\mu\varphi^\ast=(\partial_\mu+ieA_\mu)\varphi^\ast\,.\phantom{\frac12}
\end{array}\end{equation}
where $\alpha,\beta,\gamma$ are coupling constants, and $e$ is
electric change. The interaction is consistent\footnote{The concept
of consistency of interaction in the variational formalism is
reviewed in \cite{Hint}. For consistency of interactions for not
necessarily Lagrangian equations, see \cite{KLS13}. } because the
Lagrangian is invariant with respect to the usual $U(1)$ gauge
transformations. The interacting theory (\ref{AB-Pod-U}) corresponds
to the pattern of interaction for the mechanical model
(\ref{S-xy-int}) with nonzero function $U$,  which depends on the
original vector field and the complex scalar $\varphi(x)$. The
inclusion of the scalar field is essential. The Born-Infield-type
interactions, which are expressed in terms of the higher degrees of
strength tensor $F_{\mu\nu}$, are irrelevant to stability because
the energy of the model is unbounded in the linear approximation.
The model (\ref{AB-Pod-U}) describes the non-minimal couplings of
the vector field and complex scalar field, being tachyon at free
level. The $|\varphi|^4$ stabilizes the dynamics of scalar field,
while the term  $|\varphi|^2F_{\mu\nu}F^{\mu\nu}$ dynamically
generates the mass of the vector field.

The theory (\ref{AB-Pod-U}) corresponds to the model of single
higher derivative vector field $A_\mu(x)$ and scalar field
$\varphi(x)$. The field equations  read
\begin{equation}\label{}\begin{array}{c}\displaystyle
    \frac{\delta S}{\delta \varphi^\ast}=\Big(D^\mu D_\mu-\alpha+\beta|\varphi|^2+
    \frac{1}{4}\gamma F_{\rho\sigma}F^{\rho\sigma}\Big)\varphi=0\,,\quad
    \frac{\delta S}{\delta \varphi}=\Big(D^\mu D_\mu-\alpha+\beta|\varphi|^2+
    \frac{1}{4}\gamma F_{\rho\sigma}F^{\rho\sigma}\Big)\varphi^\ast=0\,,\\[5mm]\displaystyle
    \frac{\delta S}{\delta A^\mu}=\partial^\mu (G_{\mu\nu}+|\varphi|^2F_{\mu\nu})-j_{\mu}(\varphi,A)=0\,,\qquad
    \frac{\delta S}{\delta B^\mu}=\partial^\mu (F_{\mu\nu}+kG_{\mu\nu})+m^2B_{\mu}=0\,,
\end{array}\end{equation}
where $j_\mu(\varphi,A)$ denotes the scalar field charge
\begin{equation}\label{}
    j_{\mu}(\varphi,A)=ie(\varphi^\ast D_\mu \varphi - \varphi D_\mu
    \varphi^\ast)\,.
\end{equation}
From these equations, the vector $B_\mu(x)$ can be expressed
on-shell, $B_\mu\approx-\partial^\nu
(F_{\nu\mu}-k|\varphi|^2F_{\mu\nu})+kj_\mu(\varphi,A)$. Substituting
the result into the remaining equations and accounting for
$\partial_\mu j^{\mu}\approx 0$, we obtain
\begin{equation}\label{E-Pod-U}\begin{array}{c}\displaystyle
    \phantom{\frac12}\Box\partial^\nu F_{\nu\mu}+(1-k\Box)(\partial^\rho(|\varphi|^2F_{\rho\nu})-j_\nu(\varphi,A))=0\,,\phantom{\frac12}\\[5mm]\displaystyle
    \Big(D^\mu D_\mu-\alpha+\beta|\varphi|^2+\frac{1}{4}\gamma F_{\rho\sigma}F^{\rho\sigma}\Big)\varphi=0\,,\qquad
    \Big(D^\mu D_\mu-\alpha+\beta|\varphi|^2+\frac{1}{4}\gamma F_{\rho\sigma}F^{\rho\sigma}\Big)\varphi^\ast=0\,.
\end{array}\end{equation}
These equations are non-Lagrangian if $k\neq0$. The conserved tensor
of the model (\ref{AB-Pod-U}) is the canonical energy-momentum
tensor, i.e.
\begin{equation}\label{T-Pod-U}\begin{array}{c}\displaystyle
    \Theta^{\mu\nu}=G^{\mu}{}_\rho F^{\nu\rho}+G^{\nu}{}_\rho F^{\mu\rho}-
    \frac{1}{2}\eta^{\mu\nu}G_{\rho\sigma}F^{\rho\sigma}+k(G^{\mu}{}_{\rho}G^{\nu\rho}-\frac{1}{4}\eta^{\mu\nu}G_{\rho\sigma}G^{\rho\sigma})+
    m^2(B^\mu B^\nu-\frac{1}{2}\eta^{\mu\nu}B_\rho B^\rho)+\\[5mm]\displaystyle
    +|\varphi|^2(F^{\mu}{}_\rho F^{\nu\rho}-
    \frac{1}{4}\eta^{\mu\nu}F_{\rho\sigma}F^{\rho\sigma})+D^{\mu}\varphi^\ast D^\nu\varphi+
    D^{\nu}\varphi^\ast D^\mu\varphi-\eta^{\mu\nu}D^{\rho}\varphi^\ast
    D_\rho\varphi-\alpha m^2|\varphi|^2+\frac{1}{2}\beta m^2|\varphi|^4\,.
\end{array}\end{equation}
The conserved tensor of the model (\ref{E-Pod-U}) is deduced from
the expression above by expressing the auxiliary field $B_{\mu}(x)$
in terms of the derivatives of $A_{\mu}(x)$ on shell.

Consider the issue of stability of the theory (\ref{E-Pod-U}). The
model has a nonzero stationary solution,
\begin{equation}\label{vf-c}
    \phantom{frac12}A_\mu(x)=0,\qquad \varphi(x)=\varphi_0\equiv e^{i\theta}\sqrt{\frac{\alpha}{\beta}}\,,\phantom{frac12}
\end{equation}
where $\theta$ is the angle of the vacuum, and $\alpha,\beta$ are
interaction parameters. In the vicinity of this solution, the
conserved tensor (\ref{T-Pod-U}) reads:
\begin{equation}\label{T-Pod-v}\begin{array}{c}\displaystyle
    \Theta^{\mu\nu}=G^{\mu}{}_\rho F^{\nu\rho}+G^{\nu}{}_\rho F^{\mu\rho}-
    \frac{1}{2}\eta^{\mu\nu}G_{\rho\sigma}F^{\rho\sigma}+k(G^{\mu}{}_{\rho}G^{\nu\rho}-\frac{1}{4}\eta^{\mu\nu}G_{\rho\sigma}G^{\rho\sigma})+
    m^2(B^\mu B^\nu-\frac{1}{2}\eta^{\mu\nu}B_\rho B^\rho)+\\[5mm]\displaystyle
    +\frac{\gamma\alpha}{\beta}(F^{\mu}{}_\rho F^{\nu\rho}-
    \frac{1}{4}\eta^{\mu\nu}F_{\rho\sigma}F^{\rho\sigma})+\partial^{\mu}\phi^\ast \partial^\nu\phi+
    \partial^{\nu}\phi^\ast \partial^\mu\phi-\eta^{\mu\nu}\partial^\phi\phi^\ast\partial_\rho\phi+
    \alpha m^2(e^{i\theta}\phi+e^{-i\theta}\phi^\ast)^2\,,
\end{array}\end{equation}
where $\phi(x)=\varphi(x)-\varphi_0$ is the scalar field
fluctuation, and $B_\mu=\partial^\nu F_{\nu\mu}$. The $00$-component
of the conserved tensor (\ref{T-Pod-v}) reads
\begin{equation}\label{T00-Pod-v}\begin{array}{c}\displaystyle
    \Theta^{00}=G^{i}F^{i}+G^{ij}F^{ij}+\frac{\gamma\alpha}{2\beta}(F^{i}F^{i}+F^{ij}F^{ij})+
    \frac{k}{2}(G^{i}G^{i}+G^{ij}G^{ij})+\frac{1}{2}m^2(B^0B^0+B^iB^i)+
    \\[5mm]\displaystyle+\partial^{0}\phi^\ast\partial^0\phi+\partial^i\phi^\ast\partial^i\phi+\alpha m^2(e^{i\theta}\phi+e^{-i\theta}\phi^\ast)^2\,,
\end{array}\end{equation}
where $F^i=F^{0i}, G^{i}=G^{0i}$. The summation over the repeated
index $i,j=1,\ldots,d-1$ is implied. The quadratic form
$\Theta^{00}$ is positive definite if:
\begin{equation}\label{abc-Pod-cond}
    \frac{k\gamma\alpha}{\beta}-1>0\,.
\end{equation}
In this range of the coupling parameters $\alpha,\beta,\gamma$, the
interacting theory of small fluctuations in the vicinity of
stationary solution (\ref{vf-c}) is stable. We note that for the
stable interactions  the parameter $k$ should be strictly positive,
so the higher derivative theory (\ref{E-Pod-U}) is inevitably
non-Lagrangian at interacting level.

Let us discuss the dynamics of the stable theory (\ref{E-Pod-U}),
(\ref{abc-Pod-cond}). The linearization of equations of motion
(\ref{E-Pod-U}) in the vicinity of stationary solution (\ref{vf-c})
takes the following form
\begin{equation}\label{Pod-v}
    \bigg(\Box+m^2\frac{\alpha}{\beta}\bigg)\partial^\nu F_{\nu\mu}=0\,,\qquad
    \Box\phi+\alpha m^2(\phi+e^{2i\theta}\phi^\ast)=0\,,\qquad \chi=\varphi-\varphi_0\,.
\end{equation}
In the sector of vector fields, we have the usual Podolsky theory
without a resonance. The spectrum of mass of the vector field theory
includes the massless state, and the massive state with the mass
$m\sqrt{\beta/\alpha}$. This theory is stable and unitary at free
level, see in \cite{KLS}. The complex scalar field is decomposed
into two real components
$\phi^+=e^{-i\theta}\phi+e^{i\theta}\phi^\ast$ and
$\phi^-=i(e^{-i\theta}\phi-e^{i\theta}\phi^\ast)$. The scalars enjoy
the Klein-Gordon and d'Alembert equations, respectively,
\begin{equation}\label{}
    (\Box+2\alpha m^2)\phi^{+}=0\,,\qquad \Box\phi^-=0\,.
\end{equation}
The mass of the first field is $\sqrt{2\alpha} m$, while the second
one is massless. Summarizing all the above, we conclude that the
solutions of the equation (\ref{Pod-v}) transform under the unitary
representation of the Poincar\'e group. The set of
sub-representations includes the massive vector and scalar, and
massless vector and scalar. The role of the scalar field is seen
from comparison of equations (\ref{E-Pod-U}) and (\ref{Pod-v}). The
nonzero value of the scalar field $\varphi$ generates the mass term
for the vector field. The similar mechanism  is served by the Higgs
field in the standard model. With this regard, the above method of
construction of stable interactions in the higher derivative field
theories with resonance can be viewed as the Higgs-like mechanism.

Let us comment on the "Higgs mechanism" of stabilizing the higher
derivative dynamics from slightly different point of view, being
unrelated to the existence of the resonant solutions. It is a common
wisdom that the PU oscillator with non-degenerate frequency spectrum
is equivalent to the system of free harmonic oscillators. The
canonical Ostrogradski's energy of the PU action includes the
energies of these oscillators with the opposite signs. This
corresponds to the canonical energy of the first order action, being
a combination of harmonic oscillators with the alternating signs.
Corresponding Lagrangian admits the interaction vertices such that
the corresponding energy has a local minimum, so the dynamics is
stable in the vicinity of this shifted equilibrium point. The
Lagrangian reads
\begin{equation}\label{}
L=m\bigg(\frac{1}{2}(\dot{x}^2-\omega_x^2x^2)-
    \frac{1}{2}(\dot{y}^2-\omega_y^2y^2)+\frac{1}{2}(\dot{z}^2+\omega_z^2z^2)-\frac{1}{4}\alpha\omega_z^2z^4
    +\frac{1}{2}\beta z^2(\dot{y}^2-\omega_y^2y^2)\bigg)\,,
\end{equation}
where $\omega_x,\omega_y,\omega_z$ are the frequencies,
$\omega_x\neq\omega_y$, and $\alpha,\beta>0$ are the coupling
constants.  The variables $x,y$ can be thought of as the degrees of
freedom of the original PU oscillator, while $z$ can be be viewed as
the ``Higgs mode". Upon inclusion of the interaction, the
equilibrium position shifts to the point $x,y=0,z=\sqrt{1/\alpha}$.
If $\beta/\alpha>1$, the energy is positive in the vicinity of the
equilibrium position. This indicates the  stability in the vicinity
of the equilibrium. Once the frequency spectrum is non-degenerate,
the PU equations admit globally stable interactions \cite{KLS}
unlike the degenerate case, while the Higgs-like mechanism results
in the local stability in both cases.

\section{Higher derivative Yang-Mills theory}
In the previous section we considered a mechanism for including the
stable interaction in the abelian higher derivative gauge theory.
The key tool for doing that is the series of equivalent actions
(\ref{AB-Pod}) without higher derivatives involving the parameter
$k$ which does not contribute to the equations at free level. The
action (\ref{AB-Pod}) admits inclusion of $k$-dependent stable
interaction (\ref{AB-Pod-U}) such that the auxiliary fields $B$ can
be still excluded on shell. This leads to the field equations with
stable interactions, though the vertices are not Lagrangian in the
higher derivative picture. This mechanism admits, to some extent, a
non-abelian extension. At the level of the first order action
(\ref{AB-Pod-U}), the non-abelian generalisation is obvious
\begin{equation}\label{YM}
    \mathcal{L}=\frac{1}{2}(G^a_{\mu\nu}F^{a\mu\nu}+B^a_{\mu}B^\nu+\frac{1}{2}kG^a_{\mu\nu}G^{a\mu\nu})+
    \frac{1}{2}(D_\mu\varphi^a D^\mu\varphi^a+m^2(\alpha \varphi^a\varphi^a-\frac12\beta |\varphi^a\varphi^a|^2)
    +\frac{1}{4}\gamma\varphi^a\varphi^aF^b_{\mu\nu}F^{b\mu\nu}\,.
\end{equation}
Here, all the scalars and vectors take values in Lie algebra of
certain semisimple Lie group of dimension $r$,  $a=1,\ldots,r$. The
tensor $F$ denotes the Yang-Mills strength tensor of the field $A$.
The tensors $G$ is the field strength of the vector field $B$,
$G_{\mu\nu}=D_{\mu}B_\nu-D_\nu B_\mu$. The Yang-Mills covariant
derivative is $D=\partial +A$. The Lagrangian is invariant with
respect to the Yang-Mills gauge symmetry transformation. The vector
multiplet $A$ transforms as connection. The vector $B$ and scalar
$\varphi$ transform as tensors. The canonical energy is bounded for
the model (\ref{YM}) admits a local minimum. The decomposition of
Lagrangian in the vicinity of the energy minimum has the same
structure as in the abelian case (\ref{E-Pod-U}). This theory has
same stability conditions (\ref{abc-Pod-cond}). The only subtlety is
that the auxiliary field $B$ cannot be explicitly expressed from the
equations of motion unlike the abelian case. This can be done only
by perturbation theory with respect to the Yang-Mills coupling
constant. Once the auxiliary field $B$ is expressed from the
equations of motion, the non-Lagragian non-abelian higher derivative
gauge theory, which is stable and unitary.

\section{Conclusion}

In this article,  we consider the issue of stability in the class of
the higher derivative theories of derived type with a resonance. The
wave operator of the theory is the square of another lower-order
operator. We see that this class of models admits a two-parameter
series of conserved quantities. One of the entries of the series is
the canonical energy, and another one is a different integral of
motion. All  the conserved quantities are unbounded. This structure
of conserved quantities is consistent with the instability of free
model. To stabilize the dynamics of theory at the non-linear level,
the class of interactions is considered such that preserves the
selected conserved quantity of the free model. The conserved
quantity of the non-linear theory is bounded from below in the
vicinity of the equilibrium due to the interaction.  Therefore, the
fluctuations are stable in vicinity of the equilibrium . The stable
interactions are non-Lagrangian in the higher derivative equations,
but the dynamics admit the Hamiltonian formulation. The Hamiltonian
is defined by the conserved quantity of the interacting theory.
Being bounded from below, the Hamiltonian is not canonically
equivalent to any (deformation of the) Ostrogradski one.

The general scheme is illustrated by the PU oscillator of fourth
order with coinciding frequencies and by Podolsky electrodynamics
with zero mass. Both models are unstable at free level, but they can
be stabilized by an appropriate interaction. Explicit examples are
provided of the stable interactions. The Hamiltonian form of
dynamics and size of the stability island are found in the case of
the PU oscillator model.  The Hamiltonian and Poisson bracket are
explicitly derived for the model with interaction, and the
Hamiltonian is locally bounded. The PU oscillator dynamics with
stabile interactions  is also described in the original set of
variables. In the Podolsky theory, the Higgs field, being the
charged scalar, is introduced  to explicitly preserve gauge
invariance. The interacting model is a theory of higher derivative
vector field  non-minimally coupled to the charged scalar. The Higgs
field has nonzero value at the minimum of energy. The theory of
small fluctuations in the vicinity of the energy minimum has a
non-degenerate spectrum of mass. The model with non-degenerate mass
spectrum is shown to be stable and unitary.

The proposed procedure of inclusion of stable interactions seems
admitting further applications. It is consistent with the
non-abelian gauge symmetry, and it can be used in the theories of PU
type without resonance. Among the possible applications, we can
mention various higher derivative models of interest, including
gravity. The common feature is that these theories are unstable at
free and interacting level (except the class of $f(R)$-gravity
models). The introduction of appropriate set of Higgs-type fields,
can, in principle, stabilize the dynamics of the theory as we have
seen in this article.

\vspace{0.2cm} \noindent {\bf Acknowledgments.} We thank
A.A.~Sharapov for discussions on various issues addressed in this
work. The work of D.S.Kaparulin is supported by the Russian Science
Foundation grant 18- 72-10123 in association with the Lebedev
Physical Institute of RAS. S.L.~Lyakhovich is benefited from the
Tomsk State University competitiveness improvement programme.

\end{document}